\documentclass[10pt,notitlepage]{article}
\usepackage{amssymb}
\usepackage{amsmath}
\usepackage{graphicx}

\newcommand{\half}{\mbox{$\textstyle \frac{1}{2}$}}

\newcommand{\octa}{\mbox{$\textstyle \frac{1}{8}$}}

\newcommand{\bea}{\begin{eqnarray}}
\newcommand{\eea}{\end{eqnarray}}

\newcommand{\boldnabla}{\mbox{\boldmath$\nabla$}}
\newcommand{\diag}{\mathop{\mathrm{diag}}}
\newcommand{\boldeta}{\mbox{\boldmath$\eta$}}

\title{Hidden variable interpretation of spontaneous localization theory}
\author{
Daniel~J.~Bedingham\footnote{{\it Blackett Laboratory, Imperial College, London SW7 2BZ, UK.}}
\footnote{email: d.bedingham@imperial.ac.uk}
}
\date{\today}

\begin{document}
\maketitle

\begin{abstract}
The spontaneous localization theory of Ghirardi, Rimini, and Weber (GRW) is a theory in which wavepacket 
reduction is treated as a genuine physical process. Here it is shown that the mathematical formalism of 
GRW can be given an interpretation in terms of an evolving distribution of particles on configuration space 
similar to Bohmian mechanics (BM). The GRW wavefunction acts as a pilot wave for the set of particles.
In addition, a continuous stream of noisy information concerning the precise whereabouts of the particles must 
be specified. Nonlinear filtering techniques are used to determine the dynamics of the distribution of 
particles conditional on this noisy information and consistency with the GRW wavefunction dynamics is demonstrated.
Viewing this development as a hybrid BM-GRW theory, it is argued that, besides helping to clarify the 
relationship between the GRW theory and BM, its merits make it worth considering in its own right.

\end{abstract}

\section{Introduction}

The failings of standard quantum mechanics (SQM) are best exemplified with the problem of quantum measurement:
In SQM the two rules for the time-evolution of the wavefunction of a system (the Schr\"odinger equation and 
the reduction postulate) require a fundamental distinction between processes that are measurements and 
those that are not; since the concept of measurement is vague and ill-defined it follows that the theory is
vague and ill-defined.

Whilst many are content to avoid this embarrassing problem by disregarding wavepacket reduction and taking 
the view that the Schr\"odinger equation gives the complete picture, this leads unavoidably to the 
existence of macroscopic superposition states from which there is no indication of how to obtain the definite
world of our experience. The situation is concisely summed up by John Bell \cite{Bell}: 
{\it Either the wavefunction, as given by the Schr\"odinger equation, is not everything, or it is not right.}

This statement suggests two possible approaches to dealing with concerns over quantum theory. The first 
approach is to include additional state variables (hidden variables) and is well illustrated by Bohmian 
mechanics (BM) (also known as de Broglie-Bohm pilot wave theory) \cite{Bohm}. 
The second approach is to replace the Schr\"odinger equation with a more general stochastic equation capable 
of describing both unitary behavior and random wavefunction collapse events (making no reference to the 
concept of measurement). This approach is known as dynamical reduction (DR) \cite{Bass, Pear2}. (Note that in SQM we also make 
the assumption that the Schr\"odinger equation is not universally valid although the specification 
is ill-defined.)

In BM the particle positions are definite possessed properties of the system under consideration.
The wavefunction $\psi$, which satisfies the Schr\"odinger equation at all times, is viewed as a 
pilot wave whose role is to guide the trajectories of the particles. 
The flow is such that the forward equation (or continuity equation) for the probability 
distribution of particles on configuration space is identical to the equation describing the time-evolution 
of $|\psi |^2$. Therefore, in the case that the probability distribution of particles on configuration space 
is equal to $|\psi|^2$ at some initial point in time, it will also be the 
case at any later time. This condition, known as quantum equilibrium, is a subtle issue.
It is necessary in order for BM to reproduce quantum predictions (for example, it will guarantee that classically
expected particle positions will equate to quantum expectations of position operators), however, it must be
broken in general as the particles map out definite trajectories which determine such things as quantum measurement 
outcomes\footnote{Quantum equilibrium is only true in an effective sense and relies on arguments involving 
decoherence to determine the effective wavefunction to which it applies at any stage.}.

Conversely, in DR, the wavefunction describes the complete state but satisfies a stochastic generalization 
of the Schr\"odinger equation in which collapses occur randomly. This happens in such a way that for 
superpositions involving large numbers of particles the wavefunction collapse naturally occurs very rapidly 
(and with the correct quantum probability), whilst for small numbers of particles the effects are negligible.
The theory is interpreted either by treating the wavefunction as representative of matter density in the world, or 
by forming (an approximate) classical image of the world from the classical stochastic inputs to the model 
(e.g.~the discrete random collapse centers).
Here we shall focus on the spontaneous localization theory of Ghirardi, Rimini, and Weber (GRW) \cite{ghir3}, in which the 
random collapses occur in a particle position-state basis. The privileged role of the particle positions in the GRW 
theory draws parallels with BM, but overall, the two approaches are quite different.

However, the aim of this paper is to show that the mathematical formalism of GRW theory can be given an explanation 
in terms of an evolving distribution of particle positions on configuration space. On one hand this is a
novel interpretation of the theory; on the other hand this is a new BM-GRW hybrid theory employing both
additional state variables and modified wavefunction dynamics (this is the position taken in ref.~\cite{All} 
where such a BM-GRW hybrid theory is proposed). In this sense it is based on a third possibility besides the 
two given above by Bell, that the Schr\"odinger wavefunction is neither right nor is it everything.
By constructing such a theory it is possible to get the individual benefits both of BM (a clear ontological 
meaning in terms of definite particle positions making it a relatively straightforward business to interpret 
the theory), and DR (a wavefunction whose dynamics reflects the particle trajectories).

In GRW theory, $\psi$ and therefore $|\psi |^2$ exhibit stochastic behavior. If the wavefunction is to act as 
the pilot wave for a set of particles then, as with BM, we should expect that a quantum equilibrium condition
will be preserved by the dynamics. In order to show this we must demonstrate a stochastic behavior  
in the distribution of particles on configuration space equivalent to that of $|\psi|^2$. We will find that 
this is achieved by introducing a noisy `information' process relating to the true positions of the 
particles. The distribution of particles on configuration space is updated in the manner of Bayesian inference 
whereby the initial distribution represents the prior and the noisy information process constitutes an acquired 
stream of evidence. By conditioning on this noisy information we will demonstrate that
the usual particle guiding equation of BM produces a stochastic equation for the probability distribution 
of particles on configuration space which is equivalent to the GRW equation for $|\psi |^2$. In fact we will 
derive the stochastic equation for the wavefunction of GRW theory from the standard Bohmian picture $+$ noisy 
`information'.

The idea that the dynamically collapsing wavefunction can be understood in terms of a Bayesian
updating procedure has been proposed by Brody and Hughston \cite{Dorj} who considered a discrete basis 
of energy eigenstates. Starting with a DR model in which the random collapses occur in this energy state basis 
it was shown that the model could be solved by treating the terminal energy eigenvalue as a hidden variable
whose value is gradually revealed by an appropriately defined noisy information process.
Nonlinear filtering was used to determine the best estimate of the terminal energy eigenvalue (given the noisy
information) and this was shown to be equivalent to the quantum energy expectation. The same authors, along with Macrina, went on to 
apply this technology to solve a model describing the behavior of financial assets \cite{Dorj2}. (See also
\cite{Me} for use of nonlinear filtering in solving a DR model describing a simplified EPR experiment.) Here we will 
generalize this idea to cover a continuous and dynamical hidden variable process (the underlying particle dynamics 
of BM) and apply it to a realistic DR model, namely the GRW model.

The structure of the paper is as follows. In sections \ref{BM} and \ref{GRW} we briefly outline the 
mathematical formalisms of BM and the GRW theory respectively. In section \ref{NLF} we present the nonlinear
filtering problem from the innovations approach. This concerns the estimation of an unobserved process given 
observations of a related process. After demonstrating that the equations of the GRW theory can be rederived 
from a nonlinear filtering perspective we shall discuss the meaning of what have shown in section \ref{DISC}.

\section{Bohmian mechanics}
\label{BM}

Consider a quantum system describing a set of particles whose wavefunction $\psi(x_1,x_2,\cdots;t)$ satisfies the Schr\"odinger 
equation
\begin{align}
i\frac{\partial\psi}{\partial t}  =  H\psi.
\label{SE}
\end{align}
The Hamiltonian takes the form $H = -\sum_i(1/2m_i)\nabla_i^2 + V$, where $m_i$ is the mass of particle $i$ and 
$V$ is the potential. For notational simplicity we shall use the definitions 
\begin{align}
{\boldnabla} = \left(\nabla_1, \nabla_2, \cdots \right)^{\rm T};
\quad {\bf x} = (x_1, x_2, \cdots)^{\rm T}; \quad
{\bf M} = \diag(m_1, m_2,\cdots).
\end{align}
By writing the wavefunction as $\psi = R\exp \{iS\}$ we can decompose (\ref{SE}) into the two equations
\begin{align}
\frac{\partial S}{\partial t} = -\half(\boldnabla S)\cdot {\bf M}^{-1} \cdot (\boldnabla S) + V 
- \frac{(\boldnabla\cdot{\bf M}^{-1}\cdot\boldnabla R)}{2R},
\label{HJ}
\end{align}
and
\begin{align}
\frac{\partial R}{\partial t} = -(\boldnabla R)\cdot {\bf M}^{-1}\cdot (\boldnabla S) 
- \half R (\boldnabla\cdot{\bf M}^{-1}\cdot \boldnabla S).
\end{align}
Then defining $\rho({\bf x};t)=R^2 =|\psi|^2$ we can replace the second of these equations with the forward equation 
for the quantum probability distribution
\begin{align}
\frac{\partial \rho}{\partial t} 
+ \boldnabla \cdot \left(\rho {\bf M}^{-1}\cdot \boldnabla S \right)=0.
\label{fwd1}
\end{align}
Equations (\ref{HJ}) and (\ref{fwd1}) together are equivalent to the Schr\"odinger equation; the last few lines are
nothing more than a rephrasing exercise.

In BM it is assumed that a complete specification of the state of a system includes not only the 
wavefunction but also the positions of all the particles that the state is supposed to be describing.
Consider a vector set of classical particle trajectories ${\bf X}_t$ and suppose that the velocities 
of the particles are given by the guiding equation 
\begin{align}
{\bf V}_t = \frac{d{\bf X}_t}{dt} = {\bf M}^{-1}\cdot \boldnabla S({\bf X}_t).
\label{guide}
\end{align}
This equation expresses the Bohmian particle dynamics. Given this flow, it immediately follows that the probability 
distribution for the particle positions on configuration space will satisfy a classical 
forward equation precisely of the form (\ref{fwd1}). This means that if the probability distribution for the particle 
positions is equal to $\rho$ at some initial time (quantum equilibrium hypothesis) then it will be equal to $\rho$ 
at all future times (principle of equivariance). This property ensures that the (classical) statistical properties of the 
particle positions ${\bf X}_t$ will be equivalent to the quantum statistics of the position operator ${\bf x}$ (at time $t$).

Furthermore we can interpret $S$ as the action for the system of particles and (\ref{HJ}) as the Hamilton-Jacobi 
equation where quantum effects are attributed to the peculiar quantum potential:  $-(\boldnabla\cdot{\bf M}^{-1}\cdot\boldnabla R)/(2R)$. 
At this point we might wish to discard the wavefunction altogether and attempt to take a purely classical view.
However, to do this would imply that the quantum potential would have the unusual effect of causing the overall 
probability distribution of particles $\rho$ to influence the dynamics of individual particles. For this reason the role of 
$\rho$ must be elevated from epistemical to physical. The wavefunction is therefore retained and treated as a 
physical `pilot wave' guiding the flow of particles.
It should also be noted that the wavefunction has dynamical degrees of freedom such as spin (not considered here), 
that cannot be accounted for in terms of particle positions alone. These features require a physical wavefunction.

The advantages that BM has over SQM are that it is well defined and that it offers a clear interpretational 
framework; there is no doubt as to the meaning of the theory --- it concerns the motion of a set of particles.

\section{GRW theory}
\label{GRW}

Now we turn to the spontaneous localization theory of GRW \cite{ghir3} where it is assumed that the wavefunction gives a complete 
description of the state of a system. The wavefunction does not satisfy the Schr\"odinger equation. Instead it satisfies a 
more general stochastic dynamics which can be approximated by either the Sch\"odinger equation 
or quantum state reduction in situations where either of those descriptions are appropriate.

Specifically, the Schr\"odinger equation is supplemented with random spontaneous localization events. 
Consider a quantum system describing a set of distinguishable particles. (Here, we use the term `particles' not 
in the classical sense, as with BM, but simply for convenience in describing a quantum system.)
Associated with each particle is a random sequence of Poisson distributed points in time with mass-dependent frequency 
$\lambda_i=(m_i/m)\lambda$, where $\lambda$ is a reference frequency and $m$ is a reference mass. 
Whenever one of these random times is encountered the wavefunction ceases for an instant to satisfy the Schr\"odinger 
equation and undergoes a discrete change. For particle $i$ this is described by
\begin{align}
\psi(x_1,x_2,\cdots;t) \rightarrow \psi(x_1,x_2,\cdots;t+) = L(x_i-z_i) \psi(x_1,x_2,\cdots;t),
\label{GRWrule}
\end{align}
where $L$ is the localization operator. This is given by
\begin{align}
L(x_i-z_i)= \exp\left\{{-\frac{({x}_i- {z}_i)^2}{2 \sigma^2}}\right\}.
\label{collapse}
\end{align}
Here $z_i$ is a random variable representing a preferred position in space and $\sigma$ represents the width of the
(three-dimensional) Gaussian peak. The effect of the localization operator is to focus the quantum amplitude in 
configuration space about the point $x_i=z_i$. The result is a well $x_i$-localized wavefunction.

The probability distribution for the location $z_i$ is given by
\begin{align}
\mathbb{P}({z}_i\in D_i) 
\propto \int_{D_i} d{z}_i \; \langle L^2(x_i - z_i) \rangle_t,
\label{probrule}
\end{align}
for some region $D_i$ of $x_i$ values, where we have used $\langle O \rangle_t = \int d{\bf x} \psi^*({\bf x};t) O \psi({\bf x};t)$ 
to denote the quantum expectation of operator $O$. This probability rule essentially entails that the 
localization center is more likely to be where the quantum amplitude is greater.

For the model to work the average effect of localization on the wavefunction of a single particle must be mild, otherwise, quantum 
interference effects would be lost. Even with this constraint, it follows that for bulk 
superposition states 
the wavefunction can be subject to a rapid collapse.
The sheer number of particles, each subject to mild collapse effects, along with entanglements between
different particle position states, leads to an amplification mechanism. For example, if we take both $\lambda_i$ and
$\sigma$ to be small ($\sim 10^{-17}s^{-1}$ and $\sim 10^{-7}m$ respectively \cite{ghir3}) then, for an individual particle, 
localization events are rare but effective. Consequently, the chance of an individual particle undergoing a localization 
in a given small period of time can be considered to be extremely small. However, with a bulk superposition involving 
of order $10^{24}$ particles, the chance of at least one constituent particle undergoing a localization in a small time 
frame is large and the superposition as a whole is suppressed.

The GRW theory as outlined here is a discrete theory of localization events. We can take the continuum limit by
letting the frequency of localizations go to infinity such that 
\begin{align}
\lambda\rightarrow \infty \;\;\; ; \;\;\; \sigma\rightarrow \infty\;\;\; ; \;\;\; \frac{2\lambda}{\sigma^2} = g^2,
\label{limit}
\end{align}
for some constant $g$. It is in fact found that many key effects of the GRW model depend only the combination of factors 
$g$ and not separately on $\sigma$ and $\lambda$ \cite{Bass}. The result of taking this limit is the 
self contained stochastic differential equation
\begin{align}
d \psi = & - i H \psi dt  \nonumber \\ & 
-  \octa  \left\{ {\bf x}-\langle {\bf x} \rangle_t\right\} \cdot {\bf G}^2 \cdot \left\{ {\bf x}-\langle {\bf x} \rangle_t\right\} \psi dt 
+ \half \left\{ {\bf x}-\langle {\bf x} \rangle_t\right\} \cdot  {\bf G} \cdot \psi d{\bf W}_t,
\label{coreSDE}
\end{align}
where ${\bf G}=g\sqrt{{\bf M}/m}$,
and $\{{\bf W}_t\}$ is a multivariate $\mathbb{P}$-Brownian motion whose components $W_{i;t}$ satisfy $\mathbb{E}[dW_{i;t}]=0$
and $dW_{i;t}dW_{j;t} = \delta_{ij}dt$ 
($\mathbb{E}$ denotes expectation under the measure $\mathbb{P}$)\footnote{
To take the limit given by equation (\ref{limit}) we write $\lambda_i^{-1} = dt$ and $z_i=g_i^{-1}({dW'}_i/dt)$ 
where $g_i = g\sqrt{{m_i}/{m}}$; $W'_i$ is a standard (base measure) Brownian motion and ${dW'}_i/dt$ 
corresponds to white noise. The white noise scaling factor $g_i^{-1}$ will ensure normalizability of 
$\mathbb{P}$-measure probabilities (\ref{probrule}). Then from equation~(\ref{collapse}) we have
\begin{align}
L(x_i-z_i)=\exp\left\{ \frac{g_i^2 dt}{4}\left(x_i-g_i^{-1}\frac{dW'_i}{dt}\right)^2\right\}
\propto 1-\octa g_i^2 x_i^2 dt + \half g_i x_i dW'_i.
\nonumber
\end{align}
The change of measure (from base measure to $\mathbb{P}$) defined by (\ref{probrule}) enables us to specify 
a $\mathbb{P}$-Brownian motion by $dW_i=dW'_i -g_i\langle {x}_i \rangle_t dt$. Including the contributions from 
all particles and normalizing the wavefunction leads to equation (\ref{coreSDE}). A similar comparison between 
discrete and continuous models is made in Sec IV C.~of \cite{ghir2}.
\label{cont}
}.
This equation incorporates both the wavefunction dynamics and the probability rule (\ref{probrule}). 
From now on we shall use equation (\ref{coreSDE}) as our expression of the GRW theory.

Writing the wavefunction as $\psi = R\exp\{iS\}$ it is straightforward to show that (\ref{coreSDE}) is equivalent to the 
following two equations:
\begin{align}
\frac{\partial S}{\partial t} = -\half(\boldnabla S)\cdot{\bf M}^{-1}\cdot (\boldnabla S) + V 
- \frac{(\boldnabla\cdot {\bf M}^{-1}\cdot \boldnabla R)}{2R},
\label{HJ2}
\end{align}
and
\begin{align}
dR =& -(\boldnabla R)\cdot {\bf M}^{-1}\cdot (\boldnabla S) dt - \half R(\boldnabla\cdot {\bf M}^{-1}\cdot \boldnabla S) dt
\nonumber \\ &  
-\octa  \left\{ {\bf x}-\langle {\bf x} \rangle_t\right\} \cdot {\bf G}^2\cdot \left\{ {\bf x}-\langle {\bf x} \rangle_t\right\} R dt 
+ \half  \left\{ {\bf x}-\langle {\bf x} \rangle_t\right\} \cdot  {\bf G} \cdot R d{\bf W}_t .
\end{align}
Defining $\rho=R^2$ and using $d\rho = 2RdR+(dR)^2$ (from It\^o's lemma), the second of these equation can be used to derive
the forward equation for the quantum probability distribution
\begin{align}
d\rho = - \boldnabla \cdot \left(\rho {\bf M}^{-1}\cdot \boldnabla S\right)dt 
+  \rho \left\{ {\bf x}-\langle {\bf x} \rangle_t \right\} \cdot {\bf G} \cdot d{\bf W}_t.
\label{stochRho}
\end{align}
As in the previous section, equations (\ref{HJ2}) and (\ref{stochRho}) are just a way of rewriting the wavefunction
dynamics. The notable feature of this decomposition is that the Hamilton-Jacobi equation (\ref{HJ2}) remains unchanged 
from its original form (\ref{HJ}). Only the forward equation for the quantum probability distribution (\ref{stochRho}) 
includes additional stochastic terms. Note in particular that if we set $g=0$ in (\ref{stochRho}) we recover the forward equation
(\ref{fwd1}).

There are at least two existing ways to interpret the GRW theory. One is to suppose that the wavefunction is representative
of matter density in (three-dimensional) space.  For example, we could specifically define the matter density for
constituent particle $i$ as
\begin{align}
{\cal M}_i(x;t) = m_i \int d{\bf x} \; \delta(x_i-x) |\psi({\bf x};t)|^2,
\label{matter}
\end{align}
with the total matter density given by summing over the individual matter densities for each particle. 
With this definition, a superposition of two 
displaced bulk objects would correspond (at least temporarily before collapsing) to a matter distribution divided 
between the two locations. Following collapse the matter density would all be concentrated at one of the locations
(this implies that matter density is not locally conserved). Another interpretation is to suppose 
that the collapse centers $z_i$ define a discrete (classical) image of the location of matter in space and time. Note 
that these `hits' are concentrated where the quantum amplitude is greatest. The wavefunction is then an elaborate 
means of determining the likely distribution of discrete hits. (In moving to a continuous model the role of $z_i$ is 
replaced by the stochastic process $\left\{\langle {\bf x} \rangle_t + {\bf G}^{-1}\cdot (d{\bf W}_t/dt)\right\}$,
see footnote \ref{cont}.)

In the next section we will develop our BM-GRW hybrid model. (Here the interpretation will be more clear cut: as
with BM the model concerns the motion of a set of particles.) To do this we will have to understand the way
in which the stochastic equation (\ref{stochRho}) can be thought of as the classical forward equation for a 
distribution of particles on configuration space (thus ensuring, once quantum equilibrium is assumed, that the 
statistical properties of the particles reproduce the statistical properties of the wavefunction). 
It will turn out that the particles satisfy exactly the same guiding equation as for BM (\ref{guide}). The stochasticity 
of the probability density on configuration space will be understood to result from an updating procedure based on a 
continuous stream of noisy information relating to the individual particle positions. The situation corresponds to 
the problem of nonlinear filtering.

\section{Nonlinear filtering}
\label{NLF}

In this section we explain the classical nonlinear filtering problem and its relevance to GRW theory. 
A more detailed introductory account of the nonlinear filtering problem can be found in \cite{Davis}.

All stochastic processes will be defined on a fixed probability space $(\Omega, \mathbb{P}, {\cal F})$
on which there is specified a filtration $\{{\cal F}_t\}$.
We shall be concerned with some unobserved signal process $\{{\bf X}_t\}$ and a related (noisy) observation process $\{{\bf Y}_t\}$. 
The signal process $\{{\bf X}_t\}$, assumed to be adapted to ${\cal F}_t$, cannot be observed directly. 
Information about the signal process is obtained only from
the observations $\{{\bf Y}_t\}$. Given some prior distribution for ${\bf X}_0$, the nonlinear filtering problem is 
simply to determine the distribution of ${\bf X}_t$ conditional on $\{{\bf Y}_s ;  0 \leq s \leq t \}$. 

Specifically we will consider a version of the nonlinear filtering problem in which the observation process takes the form
\bea
{\bf Y}_t =  \int_0^t {\bf G} \cdot {\bf X}_s d s + {\bf B}_t,
\label{Y}
\eea
where $(\bf B_t,{\cal F}_t)$ is a standard multivariate Brownian motion process and ${\bf G}$ is a real constant diagonal matrix;
the signal process $\{{\bf X}_t\}$ takes the form
\bea
{\bf X}_t = {\bf X}_0 + \int_0^t {\bf F}({\bf X}_s) d s,
\label{X}
\eea 
where ${\bf F}$ is some vector-valued differentiable function of ${\bf X}_t$.
For simplicity we assume that $\{{\bf X}_t\}$ and $\{{\bf Y}_t\}$ are a real-valued processes; for technical
reasons we assume that $\mathbb{E}\left[\int_0^t | {\bf G} \cdot {\bf X}_s|^2 ds\right]<\infty$, and that
${\bf X}_s$ and ${\bf B}_s$ are independent of ${\bf B}_v-{\bf B}_u$ for $s<u<v$ (this is used in the result of 
Fujisaki, Kallianpur, and Kunita \cite{fuji}, see below).

With ${\cal Y}_t:= \sigma\{{\bf Y}_s ; 0 \leq s \leq t \}$ (the $\sigma$-field generated by $\{{\bf Y}_s ; 0 \leq s \leq t \}$),
our objective is to compute quantities of the form $\mathbb{E}[h({\bf X}_t)|{\cal Y}_t]$, 
i.e.~best estimates of functions $h$ of the signal at time $t$ conditional on the information contained in the noisy 
observations between times $0$ and $t$. For any process $\eta_t$ we shall use the notation 
$\langle\eta_t\rangle:=\mathbb{E}[\eta_t|{\cal Y}_t]$.

Having presented the problem in mathematical terms we proceed by defining the innovations process,
\bea
{\bf W}_t:={\bf Y}_t-  \int_0^t {\bf G}\cdot  \langle {\bf X}_s\rangle d s.
\label{inov}
\eea
It can be shown that $({\bf W}_t, {\cal Y}_t)$ is a standard multivariate Brownian motion process as follows. 
From (\ref{inov}) and  (\ref{Y}) we have
\begin{align}
\mathbb{E}[{\bf W}_t|{\cal Y}_s] = {\bf W}_s + 
\mathbb{E}\left[\left.  \int_s^t {\bf G} \cdot \left\{  {\bf X}_u-\langle{\bf X}_u\rangle\right\} du + {\bf B}_t-{\bf B}_s \right| {\cal Y}_s\right].
\end{align}
Since ${\bf B}_t-{\bf B}_s$ is independent of ${\cal Y}_s$ and has zero expectation we find that $({\bf W}_t, {\cal Y}_t)$ is a martingale:
$\mathbb{E}[{\bf W}_t|{\cal Y}_s] = {\bf W}_s$. The quadratic variation of ${\bf W}_t$ must equal the quadratic
variation in ${\bf B}_t$ (since the quadratic variation of $\int_0^t{\bf G} \cdot \left\{  {\bf X}_s-\langle{\bf X}_s\rangle\right\} ds$
is zero), therefore $({\bf W}_t, {\cal Y}_t)$ must be a standard multivariate Brownian motion.

Now we introduce the result of Fujisaki, Kallianpur, and Kunita \cite{fuji} that every square-integrable 
martingale $(m_t, {\cal Y}_t)$ has the representation
\bea
m_t = \mathbb{E}[m_0] + \int_0^t \boldeta_s \cdot d {\bf W}_s,
\label{FKK}
\eea
where $\int_0^t \mathbb{E}[\boldeta_s \cdot \boldeta_s] d s <\infty$ and $\{\boldeta_t\}$ is adapted to 
${\cal Y}_t$. With this result we will calculate $\langle h({\bf X}_t)\rangle$.
First consider a generic real-valued ${\cal F}_t$-measurable random process $\xi_t$ of the form
\begin{align}
\xi_t = \xi_0 +\int_0^t\alpha_s d s,
\label{xi}
\end{align}
with $\int_0^t\alpha_s d s$ of bounded variation. We define
\begin{align}
\mu_t:= \langle{\xi}_t\rangle - \langle{\xi}_0\rangle - \int_0^t\langle{\alpha}_s\rangle d s,
\end{align}
and with $0<s<t$ we have
\begin{align}
\mathbb{E}[\mu_t | {\cal Y}_s] &= \mu_s +
\mathbb{E}\left[\left. \langle{\xi}_t\rangle-\langle{\xi}_s\rangle 
- \int_s^t \langle{\alpha}_u\rangle du  \right| {\cal Y}_s\right]
\nonumber\\
&=\mu_s +
\mathbb{E}\left[\left. \xi_t-\xi_s
- \int_s^t \langle{\alpha}_u\rangle du  \right| {\cal Y}_s\right]
\nonumber\\
&=\mu_s +
\mathbb{E}\left[\left.
\int_s^t \left\{\alpha_u - \langle{\alpha}_u\rangle\right\} du  \right| {\cal Y}_s\right],
\end{align}
which gives the result $\mathbb{E}[\mu_t | {\cal Y}_s]=\mu_s$, i.e.~$(\mu_t, {\cal Y}_t)$ is a martingale.

Next using (\ref{FKK}) we can represent
$\mu_t$ as a stochastic integral with respect to the innovations process. We have
\begin{align}
\langle\xi_t\rangle &= \langle\xi_0\rangle +\int_0^t\langle\alpha_s \rangle d s + \mu_t
\nonumber\\
&=\langle\xi_0\rangle +\int_0^t\langle\alpha_s\rangle d s + \int_0^t \boldeta_s \cdot d{\bf W}_s,
\label{xi2}
\end{align}
for some yet-to-be-determined process $\boldeta_t$. To find $\boldeta_t$ we use the result
\begin{align}
\mathbb{E}[\xi_t {\bf Y}_t-\langle\xi_t\rangle {\bf Y}_t|{\cal Y}_s]=
\mathbb{E}\left[\left.\xi_t {\bf Y}_t-\langle \xi_t {\bf Y}_t\rangle\right|{\cal Y}_s\right]
=0,
\label{findEta}
\end{align}
for $s<t$. From equations (\ref{Y}) and (\ref{xi}) we have
\begin{align}
\xi_t {\bf Y}_t = \xi_0 {\bf Y}_0 + \int_0^t \xi_s \left\{{\bf G} \cdot {\bf X}_s ds + d{\bf B}_s\right\} + \int_0^t {\bf Y}_s \alpha_s ds,
\label{xiy}
\end{align}
and from equations (\ref{inov}) and (\ref{xi2}) we have
\begin{align}
\langle\xi_t\rangle {\bf Y}_t = \langle\xi_0\rangle {\bf Y}_0 
&+ \int_0^t \langle\xi_s\rangle\left\{{\bf G} \cdot \langle {\bf X}_s \rangle ds + d{\bf W}_s\right\} 
\nonumber\\
&+ \int_0^t {\bf Y}_s \left\{\langle\alpha_s\rangle ds+\boldeta_s \cdot d{\bf W}_s\right\} +\int_0^t\boldeta_s ds.
\label{xi2y}
\end{align}
Inserting these two expressions into (\ref{findEta}) we find that $\boldeta_t$ satisfies
\begin{align}
\boldeta_t = {\bf G} \cdot \left\{ \langle \xi_t {\bf X}_t \rangle - \langle \xi_t \rangle \langle {\bf X}_t \rangle\right\},
\end{align}
and combining this result with equation (\ref{xi2}) we have
\bea
\langle{\xi}_t\rangle = \langle{\xi}_0\rangle + \int_0^t \langle{\alpha}_s\rangle d s +  \int_0^t\left\{
\langle{\xi_s {\bf X}_s}\rangle - \langle{\xi}_s\rangle \langle{\bf X}_s\rangle \right\} \cdot {\bf G} \cdot d {\bf W}_s.
\label{NLfilter}
\eea
This is the key result for our nonlinear filtering problem. This equation describes the dynamics of the best estimate of the random 
variable $\xi_t$ conditional on the noisy information ${\cal Y}_t$. Note that when $\xi_t$ takes the form $\xi_t = h({\bf X}_t)$, 
the process $\alpha_t$ is given by $\alpha_t = {\bf F}({\bf X}_t) \cdot \boldnabla h({\bf x})|_{{\bf x}={\bf X}_t}$.

More generally we can consider the probability distribution of ${\bf X}_t$  conditional on ${\cal Y}_t$ which we represent by
$\rho({\bf x};t)$. Given that $\mathbb{E}[h({\bf X}_t)|{\cal Y}_t] = \int d{\bf x} h({\bf x}) \rho({\bf x};t)$, we 
can perform a straightforward integration by parts in equation (\ref{NLfilter}) to derive the conditional forward equation 
\begin{align}
d \rho({\bf x};t)  = -\boldnabla \cdot \left\{ \rho({\bf x};t) {\bf F}({\bf x})\right\} dt 
+ \rho({\bf x};t) \left\{{\bf x} - \langle {\bf X}_t \rangle \right\} \cdot {\bf G} \cdot d {\bf W}_t.
\label{NLrho}
\end{align}
This equation describes the time development of the probability distribution for ${\bf X}_t$ based on the fact that 
(i) ${\bf X}_t$ is a random process with dynamics described by (\ref{X}); and that (ii) the probability density is 
continually updated with noisy information concerning ${\bf X}_t$ of the form (\ref{Y}).

By now the similarity of (\ref{NLrho}) to (\ref{stochRho}) is clear. Once we interpret the signal process as the
positions of a set of particles in space we can complete the picture by choosing 
${\bf F}({\bf X}_t)={\bf M}^{-1}\cdot \boldnabla S({\bf X}_t;t)$ where $S$ satisfies the GRW equations (\ref{HJ2}) and (\ref{stochRho})
(${\bf F}({\bf X}_t)$ as defined here will be ${\cal Y}_t$-previsible).

From (\ref{X}) we then recover the guiding equation (\ref{guide}). Assuming that the quantum equilibrium 
hypothesis is satisfied at some initial time $t$ (whereby the probability distribution of particle positions on configuration space is
equal to the GRW quantum probability distribution $|\psi|^2$), we find that the quantum expectation of ${\bf x}$ is the 
same as the conditional stochastic expectation of ${\bf X}_t$
\begin{align}
\langle {\bf x} \rangle_t =
\int d{\bf x} \psi^*({\bf x};t) {\bf x} \psi({\bf x};t)=
\int d{\bf x} {\bf x} \rho({\bf x};t) = 
\langle {\bf X}_t \rangle.
\end{align}
Finally taking the matrix ${\bf G}$ of this section to be equal to the matrix ${\bf G}$ defined in the previous 
section we have shown that (\ref{NLrho}) and (\ref{stochRho}) are equivalent. Through this equivalence we identify the 
$\mathbb{P}$-Brownian motion of the GRW equation (\ref{coreSDE}) (which is related to the random localization centers) 
with the innovations process.

To summarize we can express the GRW theory in terms of a set of true particle positions satisfying a Bohmian particle guiding 
equation of the form (\ref{guide}) but dependent on an action $S$ determined from the GRW equations. 
Given an initial state of quantum equilibrium, the distribution of particle positions on configuration space will satisfy an 
equation precisely of the form (\ref{stochRho}) provided that we update the distribution with noisy information (\ref{Y}) 
concerning the precise whereabouts of the particles. This means that a state of quantum equilibrium will be sustained by the 
equations of motion. 

We could at this stage claim to have rederived the equation for the wavefunction satisfied in GRW theory 
from the standard equations of BM by introducing this noisy information process. Starting with the system of equations 
(\ref{HJ})(\ref{fwd1})(\ref{guide}), the additional noisy information forces us to replace (\ref{fwd1}) with (\ref{NLrho}). 
The result is the stochastic GRW equations for the wavefunction where the noisy information process seen as the source of 
the stochastic behavior.

\section{Discussion}
\label{DISC}

Having demonstrated that the BM-GRW theory is self-consistent, let us return to the question of why we would attempt 
to construct it in the first place. The first objective is simply to better understand the relationship between 
BM and GRW. From this point of view the BM-GRW theory acts as a stepping stone between the two. From BM-GRW we can either 
move to BM by removing the noisy information process and its effect on the wavefunction, or, move to GRW by dropping 
the hidden particle trajectories and regarding the wavefunction on its own. This helps to clarify the relationship 
between the two underlying theories. However, we will argue that the advantages of BM-GRW make it worth considering 
as a theory in its own right.

A common criticism of BM is that, whereas the wavefunction has an influence on the set of particles, the particles have 
no influence over the wavefunction. Not only does this conflict with the universal principle for laws of physics 
stating that any action is matched by a reaction, it also leads to a lot of redundancy in the wavefunction. For every
branch of the wavefunction containing the actual particle trajectories, there are countless other branches corresponding to
every other potential `world' which would have been realized had the particle positions been different. The effects of 
decoherence soon disable the influence of other branches on the particle trajectories, leaving much of the wavefunction
redundant. Nonetheless these redundant branches are an essential element of BM\footnote{
This criticism of BM has led several authors to argue that BM is little more than a version of the many-worlds 
interpretation in which the particle trajectories are a way to select one particular world. E.g., see \cite{harv}, and
for some counter arguments see \cite{val}. It has also led D\"urr, Goldstein, and Zangh\`i \cite{Durr} to suggest 
that the wavefunction should be regarded as nomological, with a role analogous to the Hamiltonian in classical mechanics.
}.

In the BM-GRW theory the particle positions do influence the wavefunction. This comes from identifying the noise in the
GRW equations for the wavefunction with the innovations process which represents information about the signal process
(the particle positions).  The influence of particles on the wavefunction is unusual in the sense that it does not result from 
a direct interaction between particles and wavefunction (as can be said for the particle guiding equation). Rather it results 
from a transfer of information from particles to wavefunction. The outcome is that the complete wavefunction continually 
reflects the true particle trajectories and the redundant branches find themselves diminished.

This in turn implies that for the BM-GRW theory it is possible to adhere to a strict form of quantum equilibrium
involving the complete wavefunction, even when describing processes such as quantum measurements. 
As discussed in the introduction, for BM, quantum 
equilibrium is only satisfied for an effective wavefunction. Although this is not necessarily a criticism, 
it does mean that a degree of judgment is required to determine what the effective wavefunction is.
The standard procedure (see \cite{Durr2}) is to first divide the total configuration into subsystem and environment 
configurations ${\bf X}_t=({\bf X}_t^{\rm S},{\bf X}_t^{\rm E})$. The effective wavefunction is then given by 
replacing the environment degrees of freedom in the complete wavefunction with their actual configurations: 
$\psi_{\rm eff}({\bf x}^S)=\psi({\bf x}^S,{\bf X}_t^{\rm E})$.
In a quantum measurement situation we can choose the macroscopic measuring device to be the environment.
The dynamics of ${\bf X}_t^{\rm E}$, given by (\ref{guide}), typically mean that the effective wavefunction does not
strictly satisfy a (subsystem) Schr\"odinger equation. In fact, the configuration of the measuring device 
(post measurement) will determine a collapse of the effective wavefunction.

In the BM-GRW theory the wavefunction actually collapses about the true particle trajectories.
(In particular, bulk superpositions such as those involving macroscopic measuring devices are naturally suppressed
by the wavefunction dynamics).
The ambiguous procedure for determining the effective wavefunction is therefore unnecessary since the complete
wavefunction $\psi$ is always representative (to the best of our knowledge) of the distribution of particles 
on configuration space.

Finally BM-GRW allows us to take a new perspective when interpreting the GRW theory. In BM we understand that the 
theory ultimately describes the behavior of a set of particles. Particles are taken to constitute the substance 
of the world. The wavefunction is seen as a real element of the theory but only insofar as it can influence the 
behavior of particles. We can carry this interpretation over to the BM-GRW theory and also use it to justify 
equation~(\ref{matter}) for the matter density in GRW.


\begin{thebibliography}{99}


\bibitem{Bell} J.~S.~Bell, {\it Speakable and unspeakable in quantum mechanics}, Cambridge (2004).


\bibitem{Bohm} D. Bohm, Phys. Rev. {\bf 85}, (1952) 166 \& 180.




\bibitem{Bass} A. Bassi \& G.C. Ghirardi, Phys. Rept. {\bf 379}, (2003) 257.

\bibitem{Pear2} P. Pearle, in: {\it Open Systems and Measurement in
Relativistic Quantum Field Theory}, H.P. Breuer \& F. Petruccione eds.,
Springer Verlag (1999).




\bibitem{ghir3} G.C Ghirardi, A. Rimini, \& T. Weber, Phys. Rev. {\bf D34}, (1986) 470.

\bibitem{All} V.~Allori {\it et al}., Brit. J. Phil. Sci., (2008) 353.



\bibitem{Dorj} D. Brody \& L. Hughston, J. Phys. A: Math. Gen. {\bf 39}, (2006) 833.


\bibitem{Dorj2} D. Brody, L. Hughston, \& A. Macrina, Int. J. Theor. Appl. Fin. {\bf 11}, (2008) 107


\bibitem{Me} D. J. Bedingham, J. Phys. A: Math. Theor. {\bf 42}, (2009) 465301.



\bibitem{ghir2} G.C. Ghirardi, P. Pearle, \& A. Rimini.  Phys. Rev. {\bf A42}, (1990) 78.




\bibitem{fuji} M. Fujisaki, G. Kallianpur, \& H. Kunita, Osaka J. Math. {\bf 9}, (1972) 19.

\bibitem{Davis} M. Davis \& S. Marcus,  in: {\it Stochastic Systems: The mathematics of filtering 
and Identification and applications}, M. Hazewinkel \& J. C. Willems eds., Reidel (1981). 



\bibitem{harv} H. R. Brown \& D. Wallace, Found. Phys. {\bf 35}, (2005), 517.


\bibitem{val} A. Valentini, in: {\it Many Worlds? Everett, Quantum Theory, and Reality}, S.~Saunders {\it et al}. eds., Oxford (2010).


\bibitem{Durr} D. D\"urr, S. Goldstein, \& N. Zangh\`i,  in: {\it Experimental Metaphysics: Quantum Mechanical Studies for 
Abner Shimony, Volume One}, R.~S.~Cohen, M.~Horne, \& J.~Stachel eds., Springer (1997).

\bibitem{Durr2}D. D\"urr, S. Goldstein, \& N. Zangh\`i, J. Stat. Phys. {\bf 67}, (1992) 843.

\end{thebibliography}
\end{document}